\documentclass[prc,twocolumn,floatfix,groupedaddress,showpacs] {revtex4}
\usepackage{graphicx}% Include figure files
\graphicspath{{./no_wf/}{./error_and_variance/}{./plot18Gcrit/}{./cofvariation/}{./excited18/}{./cont_correction/}}
\usepackage{dcolumn}%Align table columns on decimal point
\usepackage{mathrsfs}
\usepackage{bm}% bold math
\usepackage{dsfont}
\usepackage{dcolumn}% Align table columns on decimal point
\usepackage[usenames]{color}
\usepackage{amssymb}
\usepackage{amsmath}
\usepackage{makeidx}
\usepackage{enumitem}
\newcommand\textequal{%
 \rule[.4ex]{4pt}{0.4pt}\llap{\rule[.7ex]{4pt}{0.4pt}}}
 \newcommand{\sdots}{...}
 \newcommand{\shorteq}{\textequal\,}
 
\DeclareMathAlphabet{\mathpzc}{OT1}{pzc}{m}{it} %define calligraphic font for small
\usepackage{color}
 
\begin{document}

\title{Study of nuclear pairing with Configuration-Space Monte-Carlo approach}
\author{Mark Lingle}
\author{Alexander Volya}
%\email{avolya@fsu.edu} 
\affiliation{Department of Physics, Florida State University, Tallahassee, FL 32306, USA} 

\date{\today}
\begin{abstract}
Pairing correlations in nuclei play a decisive role in determining nuclear drip-lines, binding energies, and many collective properties. 
In this work a new Configuration-Space Monte-Carlo (CSMC) method for treating nuclear pairing correlations is developed, implemented, and 
demonstrated.
In CSMC the Hamiltonian matrix is stochastically generated in Krylov subspace, resulting in the Monte-Carlo version of Lanczos-like diagonalization.  
The advantages of  this approach over other techniques are discussed; 
the absence of the fermionic sign problem, probabilistic interpretation of quantum-mechanical amplitudes, 
and ability to handle truly large-scale problems with defined precision and error control, are noteworthy merits of CSMC. 
The features of our CSMC approach are shown using models and realistic examples.
Special attention is given to difficult limits: situations with non-constant pairing strengths, cases with nearly degenerate excited states, 
limits when pairing correlations in finite systems are weak, and problems when the relevant configuration space is large.  
\end{abstract}
%21.60.Cs Shell model, add more
%02.70.Ss	Quantum Monte Carlo methods
%21.60.Ka Monte Carlo models
\maketitle

\section{Introduction}
\label{intro}

Pairing correlations are a salient component of the nuclear many-body dynamics which has a profound impact on most of the nuclear properties, 
and on the nuclear landscape in general.  The recently published volume {\em Fifty Years of Nuclear BCS} \cite{BCSBOOK} offers a unique overview of more than fifty years of research in this area. 
Advances in experimental techniques and emergence of many new facilities made it possible to explore nuclear systems at the edge of stability. 
Interest in pairing is reinvigorated by extraordinary effects observed recently in near-drip line nuclei.
This includes the so-called nuclear halo effect in neutron rich nuclei, such as the case of {\em borromean} nucleus $^{11}$Li 
which is bound only due to the specifics of the pairing dynamics of the two neutrons above the $^9$Li core.  
Recent observations of {\em di-neutron} decay \cite{Spyrou:2012} and other remarkable manifestations of pairing,  seen in structure 
and reactions with exotic nuclei \cite{Hoffman:2011, Kohley:2013}, all encourage theoretical effort to be continued.

% definitions of the pairing Hamiltonian
The pairing interaction involves pairs of time-conjugate single-particle states.   We use $p_{k}^{\dagger}$, $p_{k},$ and $\hat{n}_{k}$ to denote the corresponding pair creation, pair annihilation, and number-of-pairs operators; index $k$ is used to identify various 
distinct pair-states in the system.
The pairing Hamiltonian of interest is defined as:
\begin{equation}
H=2\sum_{k}\epsilon_{k}\hat{n}_{k}-\sum_{k,k'}{G_{kk'}}p_{k}^{\dagger}p_{k'},
\label{eq:ph}
\end{equation}
Here $\epsilon_k$ are the single-particle energies and $G_{kk'}$ are the matrix elements of pairing interaction. 
In this work we limit our discussion to fully paired systems of $n$ pairs in $\omega$ pair-states, which corresponds to $2n$ fermions 
within the 
total particle capacity $2\omega$ of the valence space. Working under the assumption of a fully paired state 
is completely general: any unpaired nucleons 
remain untouched by the Hamiltonian (\ref{eq:ph}); these nucleons effectively block some part of the valence space so that the 
problem is then reduced to a fully paired state in a reduced space. 
Thus, the configuration space of interest spans over all $\omega$ choose $n$ basis states
\begin{equation}
|{\bf n}\rangle = |n_{1},n_{2},\dots n_{\omega}\rangle.
\label{eq:basis}
\end{equation}
Here we use occupation representation where for each pair-state $n_{k}=\langle{\bf n}|\hat{n}_k|{\bf n}\rangle = 1$ or $0$ depending on whether 
the pair-state is occupied or not. Clearly, the total number of pairs $n=\sum_k n_k.$   
Any state can be represented as a linear combination of the basis states,
$|\Phi\rangle=\sum_{\bf n} \langle {\bf n} | \Phi\rangle\, |{\bf n}\rangle.$

%end pairing Hamiltonian definition

Pairing correlations have been traditionally explored with the help of the BCS theory of superconductivity \cite{Bardeen:1957}. 
This variational technique, which is formally exact in thermodynamic limit, is very well integrated into more general mean-field approaches and into techniques beyond mean-field. 
Starting from pioneering works \cite{Bohr:1958,Belyaev:1959}, the BCS theory has been applied in nuclear physics with great success. However, 
non-conservation of the particle number and difficulty in handling limits where pairing is weak as compared to the characteristic 
mean-field single-particle level spacing, have proven to be significant drawbacks in applications of BCS to 
finite nuclear systems \cite{Kerman:1961,BCSBOOK,Dukelsky:2003,Zelevinsky:2003,Zelevinsky:2004}. 
Over the years a number of remedies have been proposed to overcome these drawbacks. For example, the issue of the particle number 
non-conservation has been 
addressed with a variety of techniques proposed in Refs.~\cite{Pang:1972, Zeng:1983,Dang:1966,Dang:1966_2,Lipkin:1960,Nogami:1964,Pan:1998,Volya:2012}. 

With theoretical and computational advances a growing number of pairing problems in  mesoscopic systems, such as atomic nuclei, 
can be treated exactly; thus avoiding the BCS and its drawbacks.  
There are several major groups of exact methods.
Symmetry-based algebraic methods were introduced by Racah \cite{Racah:1942,Racah:1942_2,Racah:1943} even before the BCS theory.
These methods found wide applicability both independently ~\cite{Hecht:1965,Ginocchio:1965,Auerbach:1966,Kota:2006,Dukelsky:2006} and as components of other techniques~\cite{Brown:2002,Chen:2014}.

Presented more than 40 years ago by Richardson 
\cite{Richards1:1967,Richards.Rw:1966,Richards:1967,Richardson:1964,Richards.Rw:1965}, an exact solution that reduces the pairing eivenvalue 
problem to a set of non-linear equations have been successfully generalized and applied \cite{cooper56,Dukelsky:2004,Pittel:2006,Dukelsky:2006r} 
in multiple situations. Some generalizations and interpretations, such as those related to electrostatic analogies \cite{Dukelsky:2002}, are of particular theoretical interest \cite{Dukelsky:2004}.

Computational advances and iterative sparse matrix diagonalization algorithms allowed for direct diagonalization methods to emerge as extremely 
simple, stable, and robust alternatives \cite{Whitehead:1972,Volya:2001PLB,Sumaryada:2007}.  Nevertheless, the dimension of the Hamiltonian matrix in  the relevant basis space grows exponentially with the number of pairs, eventually rendering these methods computationally impractical especially for model spaces required for problems with continuum of scattered states. 
Mote Carlo approach, which is the main subject of this work, so far appears to provide the only reasonable technique that overcomes, in a controlled way,  the exponentially growing computational difficulty.   
There exist numerous variations of the Monte Carlo approach, varying in philosophy and implementation. Many of these methods can be found in 
the textbook \cite{Landau:2005qe}. The Shell Model Monte-Carlo \cite{Koonin:1997,Koonin:1997_1} and Quantum Monte-Carlo involving
variational, auxiliary-field and Green's function versions (see review \cite{Pieper:2001}) are among well-known successful examples used in 
low energy nuclear physics.  Random Monte-Carlo sampling, either for variational purposes or in order to evaluate multi-dimensional integrals, such as those emerging in Hubbard-Stratanovich transformation is at the center of these techniques.

% pairing monte-carlo

The pairing 
Hamiltonian~(\ref{eq:ph}) is equivalent to a Hamiltonian
describing $\omega$ spin-$1/2$ particles, where one can assume $n_k=1$ and 0 for up and down spin orientations, respectively, 
\cite{Volya:2001PLB,Zelevinsky:2003,Sumaryada:2007}. 
This offers opportunities 
for a broad class of Monte-Carlo methods known in spin systems \cite{Landau:2005qe} to be applied. 
The idea of using the connection between spin physics in condensed matter and 
quasispin in pairing problems  
was originally explored by Cerf and Martin \cite{Cerf:1993pb,CERF:1993bs}. 
In their approach the ground state $|\Psi_0\rangle$ is found as an asymptotic state that emerges from an arbitrary initial state $\Phi$ as
a result of evolution along the imaginary time:
\begin{equation} 
|\Phi\rangle \simeq e^{-\tau E_{0} } \langle \Psi_{0} |\Phi\rangle |\Psi_{0} \rangle, \quad {\rm as} \quad \tau \rightarrow \infty.
\label{eq:eproj}
\end{equation}
In order to propagate the initial state along the imaginary time, Cerf and Martin proposed breaking the Hamiltonian into the two non-commuting parts $H_{1}$ and $H_{2},$ corresponding to one-body and two-body terms. Then 
the propagation can be done in small steps $\Delta\tau$ using  Trotter-Suzuki operator decomposition \cite{Trotter:1959t,Suzuki:1990s},
\begin{equation}
e^{-\Delta\tau(H_{1}+H_{2})} = e^{-\Delta\tau \frac{H_{1}}{2}}e^{-\Delta\tau H_{2}}e^{-\Delta\tau \frac{H_{1}}{2}}+ O\left(\Delta\tau^{3}\right).
\end{equation}
The principal advantage of the technique is that $H_{1}$ is diagonal in the basis states $|{\bf n}\rangle$ and the corresponding exponent can be easily evaluated. Assuming a constant pairing, where $G_{kk'}\equiv G,$ Cerf and Martin proposed to evaluate $e^{-\Delta\tau H_{2}}$ stochastically by breaking the exponent into a Taylor series and taking advantage of the fact that for constant pairing strength the probability of a walk in configuration space to have a given number of steps is exactly Poissonian. Given that for constant pairing $H_2$ can be diagonalized analytically using quasispin algebra, it may be possible to use the Trotter-Suzuki propagation without involving Monte-Carlo.  

With some degree of success, the method of Cerf and Martin was picked up recently by other research groups 
\cite{Mukherjee:2011la,Alhassid:2012amno}.
The algorithm has so far been applied mainly to problems with constant pairing matrix elements. This apparent limitation is not well addressed in the literature, but appears to be related to unknown quality and reliability of the stochastic evaluation of exponential operator of the two-body interaction with non-constant matrix elements.

In the Configuration-Space Monte-Carlo (CSMC) 
algorithm presented in this work we completely avoid the imaginary time evolution and Trotter-Suzuki 
decomposition; instead, using a stochastic process of nucleon-pair diffusion through the configuration space, 
we built a Krylov subspace which contains 
the set of lowest eigenvalues.  The resulting algorithm can be seen as a Monte-Carlo version of the well-known Lanczos algorithm. This class of
algorithms are often referred to as projector algorithms \cite{Landau:2005qe} since  repeated application of the Hamiltonian operator 
to a random state eventually amounts to the ground state being projected out. Excited states can be obtained as well by storing the wave functions 
and by enforcing orthogonality.

\section{Configuration Space Monte-Carlo}
In this section we present the Configuration Space Monte-Carlo method. It should be mentioned that the method is generic, 
and is not limited to pairing, 
but the specifics of the pairing Hamiltonian offer some big advantages, which is discussed in Sec.~\ref{sec:features} and demonstrated in 
Sec.~\ref{sec:pmc}.  

\subsection{CSMC formalism}
Let us consider a sequence of states 
\begin{equation}
|\Phi_{L} \rangle \equiv V^{L} |\Phi_0 \rangle
\label{eq:power}
\end{equation}
which is generated by a repeated application of the Hamiltonian $H=-V$ onto a random initial vector $|\Phi_0 \rangle$. These states span over 
the Krylov subspace. Eigenvalues of the Hamiltonian matrix in this subspace converge, after enough iterations, to the eigenvalues 
(greatest in absolute value) of the Hamiltonian
in the entire space. Since we are interested in the lowest, most negative, states it is convenient to carry out this discussion using $V=-H.$
The repeated application of the operator $V$ can be written as a summation over all possible $L+1$ intermediate states which are given by the sets  
$\{{\bf n}\}_L\equiv\{{\bf n}_0, {\bf n}_1,\dots {\bf n}_{L}\},$
\begin{equation}
| \Phi_L\rangle= 
\sum_{\{{\bf n}\}_L} 
 | {\bf n}_L \rangle  A\left (\{{\bf n}\}_L \right), 
 \label{eq:sum}
\end{equation}
where the amplitude is 
\begin{multline}
 A\left (\{{\bf n}\}_L \right) \equiv 
\langle {\bf n}_{L} | V | {\bf n}_{L-1} \rangle 
\langle {\bf n}_{L-1} | V | {\bf n}_{L-2} \rangle \\
\dots 
\langle {\bf n}_{1} | V | {\bf n}_{0} \rangle  \langle {\bf n}_{0} | \Phi_{0} \rangle. 
\label{eq:amp}
\end{multline}
One advantage of evaluating powers of the Hamiltonian operator is that the summation in Eq.~(\ref{eq:sum}) is restricted to 
all possible paths ${\bf n}_{0}\rightarrow{\bf n}_{1} \dots \rightarrow {\bf n}_{L}$ where each consecutive configuration is connected to the previous one by the matrix element of the interaction $V.$ 
Therefore, in what follows  $\{{\bf n}\}_L$ denotes a connected $L$-step long path  
${\bf n}_{0}\rightarrow{\bf n}_{1} \dots \rightarrow {\bf n}_{L}.$

The path summation can be performed using Monte-Carlo sampling. To be more specific, if one generates $N$ paths 
$\{{\bf n}\}^{(s)}_L\equiv 
{\bf n}_{0}^{(s)}\rightarrow{\bf n}_{1}^{(s)} \dots \rightarrow {\bf n}_{L}^{(s)},$ labeled here with superscript $s=1\dots N,$
then 
\begin{equation}
| \Phi_L\rangle\approx 
\frac{1}{N}
\sum_{s=1}^N 
 | {\bf n}_L^{(s)} \rangle \,
B (\{{\bf n}\}_L^{(s)} )
\label{eq:mc}
\end{equation}
where 
\begin{equation}
B (\{{\bf n}\}_L^{(s)} )\equiv\frac{A(\{{\bf n}\}^{(s)}_L)}{ {\cal P}(\{{\bf n}\}^{(s)}_L)}
\label{eq:bb}
\end{equation}
is the amplitude for the $s$-th random path  weighted by the inverse of  the probability to generate this path  ${ {\cal P}(\{{\bf n}\}^{(s)}_L)}$. 
In applications where sampling is done with uniform 
probability ${\cal P}(\{{\bf n}\}^{(s)}_L)\equiv {\cal P}$, the common term $1/{\cal P}$ is just the total number of 
all possible paths ${\{{\bf n}\}_L}$ which equals to the number of terms in the sum in Eq.~(\ref{eq:sum}).

Each sampling path can be generated as a random walk.
%Suppose after $L$ steps ${\bf n}_{0} \dots \rightarrow {\bf n}_{L}$ the position in configuration space is ${\bf n}_{L},$ the amplitude  is 
%$  A\left (\{{\bf n}\}_L \right),$ and the probability to generate this walk among all $L$-step walks is 
%${\cal P}\left (\{{\bf n}\}_L \right).$
The amplitude in Eq.~(\ref{eq:amp}) is subject to the recursion relation 
\begin{equation}
A\left (\{{\bf n}\}_{L+1} \right)=\langle {\bf n}_{L+1} | V | {\bf n}_{L} \rangle\,A\left (\{{\bf n}\}_L \right),
\label{eq:rec}
\end{equation}
where $ A\left (\{{\bf n}\}_0 \right)=\langle {\bf n}_{0} | \Phi_{0} \rangle. $
Similarly, the probability for a path is the product of probabilities for each step. Therefore, starting from the probability to pick the first configuration 
${\cal P}({\bf n}_{0})\equiv {\cal P}\left (\{{\bf n}\}_0 \right),$ the probability for the entire path is generated recursively as  
\begin{equation}
{\cal P}\left (\{{\bf n}\}_{L+1} \right)={\cal P}({\bf n}_{L} \rightarrow {\bf n}_{L+1})\,{\cal P}\left (\{{\bf n}\}_L \right);
\label{eq:recp}
\end{equation}
here ${\cal P}({\bf n}_{0} \rightarrow {\bf n}_{L+1})$ is the conditional probability to move  to configuration ${\bf n}_{L+1}$ given the 
current position 
at ${\bf n}_{L}.$ 
Thus, while going along a random path the  coefficients $B$ are generated recursively, 
\begin{equation}
B\left (\{{\bf n}\}_0 \right)=\frac{\langle {\bf n}_{0} | \Phi_{0} \rangle}{{\cal P}({\bf n}_{0})} \quad \text{and}
\label{eq:bag1}
\end{equation}
\begin{equation}
B\left (\{{\bf n}\}_{L} \right)=\frac{\langle {\bf n}_{L} | V | {\bf n}_{L-1} \rangle}{{\cal P}\left({\bf n}_{L-1} \rightarrow {\bf n}_{L} \right)}\,B\left (\{{\bf n}\}_{L-1} \right).
\label{eq:bag2}
\end{equation}
The probability distribution for selecting an initial position in configuration space  ${{\cal P}({\bf n}_{0})} $ 
and the distribution of conditional probabilities 
${{\cal P}\left({\bf n}_{L-1} \rightarrow {\bf n}_{L} \right)}$
describing the direction in which each next random step  is to be taken, are both arbitrary user-supplied functions. Strategies for selecting these functions are discussed in what follows. 

It is important that the probability of taking a certain step depends only on the current position and not on the preceding history, therefore the process represents
a Markov chain \cite{Landau:2005qe}. 
The computational implementation of the Markov Chain Monte-Carlo methods is a well studied subject; see Ref.~\cite{Landau:2005qe} and
references therein. 

The Configuration Space Monte-Carlo approach, defined by Eq.~(\ref{eq:mc}), is implemented using 
an ensemble of $N$ ``walkers'' starting from configurations  ${\bf n}_{0};$ the initial configurations are generated with the probability distribution ${\cal P}({\bf n}_{0}).$ Then each walker independently 
takes $L$ random steps; the probability distribution ${\cal P}\left({\bf n}_{L} \rightarrow {\bf n}_{L+1}\right)$ is used to generate steps.  
We envision that each walker carries a ``bag''  $B$ that is initialized and modified along the path following Eqs. (\ref{eq:bag1}) and (\ref{eq:bag2}). 
Contributions from the bags of all walkers arriving to a given configuration 
${\bf n}_{L}$ comprise the component $\langle {\bf n}_{L}|\Phi_L\rangle$ as shown by Eq. (\ref{eq:mc}).

As the most straightforward application of the method, one could assume the probabilities for steps in all ``directions'' to be equal,  
then the conditional probability 
${\cal P}\left({\bf n} \rightarrow {\bf n}' \right)$ depends only on the initial configuration ${\bf n}$ and the inverse of it equals to the number of 
configurations
connected to ${\bf n}.$ In most cases the number of connected configurations is the same for all states, 
which makes the conditional probability for each step being an absolute constant, i.e., independent of initial and final positions. 
For example, for any paired configuration with $n$ pairs and $\omega$ pair-spaces the pairing Hamiltonian 
can generally move one of the $n$ pairs onto one of the $\omega-n+1$ unoccupied pair-states 
(this includes diagonal move back to the same pair-state). 
Thus, in the pairing case, for equiprobable steps the conditional probability becomes a configuration-independent constant
 ${\cal P}\left({\bf n} \rightarrow {\bf n}' \right)=(n(\omega-n+1))^{-1}$ and the resulting random paths are all generated with equal probability. 
This amounts to uniform Monte-Carlo sampling of terms in sum (\ref{eq:sum}).

\subsection{Importance sampling\label{sec:isampling}}
Uniform sampling 
is convenient and effective when contributions from most paths are nearly equal; constant-strength pairing Hamiltonian 
discussed by Cerf and Martin in Refs. \cite{Cerf:1993pb,CERF:1993bs} is a good example of this situation.  However, sampling uniformly can 
be extremely ineffective if certain amplitudes  $A\left (\{{\bf n}\}_{L} \right)$ are very small or equal to zero;
importance sampling can be introduced as a remedy. 
In the CSMC the contributions from different sampling paths can be made comparable in magnitude if steps are generated with probabilities proportional to the 
magnitude of the corresponding matrix elements,
\begin{equation}
{\cal P}\left({\bf n}_{L-1} \rightarrow {\bf n}_{L} \right)\propto |\langle {\bf n}_{L} | V | {\bf n}_{L-1} \rangle|.
\label{eq:importance}
\end{equation}
This way the scaling factor in Eq.~(\ref{eq:bag2})  would not depend on the direction of the step.  
It should be emphasized, that satisfying the proportionality 
(\ref{eq:importance}) exactly, which may be computationally expensive, is not necessary.  Any probability distribution that in some general way follows the distribution of the matrix elements is sufficient.

The approach described here, referred to as Configuration Space Monte Carlo, allows one to build  stochastically the Krylov subspace 
and find the eigenstates and eigenvalues of the Hamiltonian using steps similar to those in Lanczos approach. 
Clearly, the method is applicable to any Hamiltonian; however, different signs of matrix elements $\langle {\bf n}_{L+1} | V | {\bf n}_{L} \rangle$ 
and thus different signs of the amplitudes can lead to poorly convergent sums. This issue, commonly known as the Monte-Carlo sign problem, is not present in applications of the CSMC to pairing problems that are discussed next.

\subsection{Features of the pairing Hamiltonian\label{sec:features}}
Let us summarize some of the important features of the pairing problem that boost the effectiveness of the CSMC method. 
\begin{description}[style=unboxed,leftmargin=0.0cm]
\item[(i)] For fully paired systems the diagonal pairing matrix elements $G_{kk}$ are equivalent to the single-particle energies.  Thus, by redefining 
the diagonal pairing matrix elements as $G_{kk}\rightarrow G_{kk}-\epsilon_k/2,$ the pairing Hamiltonian can be written in the following form
\begin{equation}
V \equiv -H,\quad\text{where}\quad V=\sum_{k,k'}{G_{kk'}}p_{k}^{\dagger}p_{k'}.
\label{eq:ph2}
\end{equation}
\item[(ii)] The pairing interaction is attractive. Therefore, with the proper choices of phases all off-diagonal matrix elements $G_{kk'}$ can be made 
non-negative. Without any loss of generality the single-particle energies can be measured relative to some chemical potential $\mu;$ and the constant $\mu$ can be selected so that all diagonal many-body matrix elements $\langle {\bf n} | V | {\bf n} \rangle$  are also positive. 
\item[(iii)] The nucleon pairs are the only degrees of freedom, and the entire dynamics is represented by the  ``hopping'' of the pairs between available 
states.
Each pair hopping leads to a step in configuration space where 
\begin{multline*}
\langle {\bf n}' | V | {\bf n} \rangle = 
\langle \sdots n_{k}\shorteq1,\sdots n_{k'}\shorteq0, \sdots | V | \dots n_{k}\shorteq0,\sdots n_{k'}\shorteq1,\sdots \rangle  \\
= G_{kk'}\ge 0$ if $k\ne k',
\end{multline*}
and  $\langle {\bf n}| V | {\bf n} \rangle=\sum_k G_{kk} n_k>0$ for the diagonal. For any initial configuration there are $n(\omega-n+1)$ 
different final configurations that can be reached in one step. 
\item[(iv)] Given that the matrix elements of $V$ are all positive, the Monte-Carlo sign problem does not appear. 
\item[(v)] Positive matrix elements of $V$ imply that if in the initial wave function $\Phi_0$ all components $\langle {\bf n} | \Phi_0\rangle\ge0,$ 
which can always be accomplished by defining phases of the basis states $|{\bf n}\rangle,$ then all components of any $\Phi_L$ are non-negative,
that is, $\langle {\bf n} | \Phi_L\rangle\ge0$ for any $L$ and any  ${\bf n}.$ This also allows one to introduce a linear  ${\cal L}_1$ norm 
\begin{equation}
||\Phi_L||\equiv  \sum_{\bf n} \langle {\bf n} | \Phi_L\rangle.
\end{equation}
\item[(vi)] Asymptotically, as ${L \rightarrow \infty},$
 \begin{equation} 
 |\Phi_L\rangle \simeq  (-E_{0})^L \langle \Psi_{0} |\Phi_0\rangle |\Psi_{0} \rangle,
\label{eq:gsproj}
\end{equation}
where $|\Psi_{0} \rangle$ is the ground state wave function and  $E_0$ is the ground state energy. Since $E_0<0,$ in fact for the negative-definite 
Hamiltonian, all eigenvalues are negative, and the phase of the ground state wave function can be selected so that 
$\langle \Psi_{0} |\Phi_0\rangle>0,$ and all components of the ground state wave function are also non-negative:  $\langle {\bf n} | \Psi_0\rangle\ge0.$ 
\item[(vii)] Given that all many-body states that span the Krylov subspace have positive-definite amplitudes relative to the basis states 
$|{\bf n}\rangle$, these
amplitudes can be treated as probabilities. Therefore 
in the ideal limit of the importance sampling Monte-Carlo, when Eq.~(\ref{eq:importance}) is satisfied exactly, all walkers' bags are equal. 
In this limit the number of walkers arriving to a certain many-body 
configuration  ${\bf n}$ on step $L$ is proportional to $\langle {\bf n} | \Phi_L\rangle.$ Linear norm can be used to normalize the wave function. 
\end{description}

\section{Nuclear Pairing with Configuration Space Monte Carlo\label{sec:pmc}}
In what follows we demonstrate the  CSMC applications to pairing problems. We organize our presentation by progressing from 
simple to more elaborate applications, we use model and realistic examples to highlight the CSMC and to address technical details.

In the following subsections A-D
we consider a system consisting of $\omega$ double-degenerate 
equally-spaced single particle orbitals, $\epsilon_k=\epsilon\, k$ with $k=0,1\dots \omega-1,$ where the single-particle level spacing $\epsilon$ defines the unit of energy. 
This system, often referred to as a picket-fence or ladder model, is commonly used for testing various approaches to 
pairing~\cite{Sumaryada:2007}. The model has a minimal symmetry, time-reversal only, making it the most computationally challenging one. 
%We consider a half-occupied system, assuming that the number of pair-states $\omega$ is even and the number of pairs $n=\omega/2.$
%The energy of the single particle levels in the picket fence model are chosen symmetrically about zero which puts a fermi and chemical potential 
%this half occupied system exactly at zero. 
For the set of studies using the ladder model we assume constant pairing interaction, where  $G_{kk'}= G$ in Eq.~(\ref{eq:ph}). 
This choice is not related to any limitation, this  
merely minimizes the number of parameters and is convenient for comparison 
with numerous previous studies where the same model was used. 
For the half-occupied ladder model the critical BCS strength is approximately $G_{\rm cr}=\epsilon/\ln(2\omega),$ see Ref.~\cite{Sumaryada:2007}. 
%It is particularly interesting to consider a range of pairing strengths near this value since this is the area where methods other
%than BCS-based are needed.  

Realistic examples with non-constant pairing strength and large scale application are discussed in subsections E and F. 

\subsection{Linear norm\label{ssec:simple}}
We start with a very simple and 
quick technique that can be used to determine the ground state energy and requires no storage for wave functions. 
Despite certain limitations, the method is  elegant, simple in implementation, very computationally efficient, and is an important 
component in the general approach. 

In the implementation of the CSMC through multiple walkers in configuration space the construction of the 
wave function $|\Phi_L\rangle$ is the most challenging task because, according to Eq.~(\ref{eq:mc}), it requires organizing 
walkers based on their arrival locations. Even if performed in parallel, this is still a daunting task when the number of contributing configurations 
becomes large.    
As we show next, for certain observables, and for the ground state energy in particular, this task can be avoided; 
thanks to the properties of pairing interaction outlined in Sec.~\ref{sec:features}.

According to Eq. (\ref{eq:mc}) the average of all bags gives the linear norm (referred to as ${\cal L}_1$) of the wave function
\begin{equation}
\frac{1}{N}
\sum_{s=1}^N 
B (\{{\bf n}\}_L^{(s)} ) \approx
 \sum_{\bf n}  \langle {\bf n} | \Phi_L\rangle \equiv ||\Phi_L||;
 \label{eq:lnorm}
\end{equation} 
computing the bag average is a simple and fast operation.

Following Eq.~(\ref{eq:gsproj}), the bag averages for two consecutive values of $L$ as ${L \rightarrow \infty}$ give an estimate for the ground state energy as  
\begin{equation}
E_0\simeq  E_0(L)\equiv- \frac{\sum_{{\bf n}} \langle {\bf n} | \Phi_{L+1}\rangle}{\sum_{\bf n} \langle {\bf n} | \Phi_L\rangle}=-\frac{||\Phi_{L+1}||}{||\Phi_L||}.
%=\frac{||H^{L+1}\Phi_0||}{||H^L\Phi_0||}.
\label{eq:HLproj}
\end{equation}
Clearly, any procedure on the Hamiltonian leading to the ground state can be subjected to the linear norm. For example, using projection (\ref{eq:eproj}) one could evaluate energy in the limit $\tau\rightarrow\infty$ as 
\begin{equation}
E_0\simeq E_0(\tau)\equiv\frac{||H e^{-\tau H} \Phi_{0}||}{||e^{-\tau H}\Phi_0||}, 
\label{eq:expproj}
\end{equation}
where exponents are evaluated in CSMC approach using Taylor series
\begin{equation}
||e^{-\tau H} \Phi_{0}||=\sum_{L=0}^\infty \frac{\tau^L}{L!}\, || \Phi_L||.
\label{eq:exptaylor}
\end{equation}
Obviously, it is possible to compute the linear norm for any operator 
$
||O \Phi || \equiv \sum_{\bf n}  \langle {\bf n} |O| \Phi\rangle, 
$
but unfortunately in most situations this linear norm does not have a transparent physical meaning. 

\begin{figure*}
\centering
\includegraphics[width=160mm]{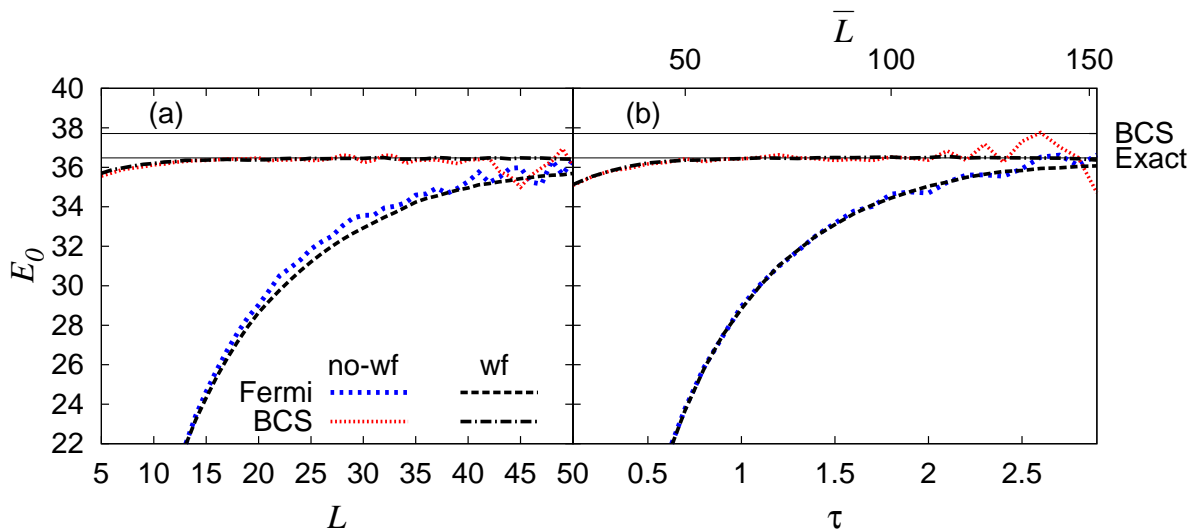}
\caption{(Color online) The half-occupied ladder model with $\omega=18,$ $n=9$ and $G=1$ is used to show the convergence of ground state energy using an
approach based on the linear norm. Left panel (a) shows the method based on ground state projection with power law Eq.~(\ref{eq:HLproj}), and 
right panel (b) shows projection using imaginary time evolution Eq.~(\ref{eq:expproj}). In both panels BCS and exact values of energy are shown 
with horizontal grid lines; the four curves represent two initial states and two methods: with and without wave functions being built. 
 }
\label{fig:no_wf_two}
\end{figure*}

This quick linear-norm-based technique for evaluating the ground state energy within CSMC algorithm
is illustrated in Fig. \ref{fig:no_wf_two},  and is compared to the general approach discussed in the following subsection.  
A ladder model with $\omega=18$ levels and $n=9$ nucleon pairs, where $G=1,$ is used in this example.
Two different starting wave functions $|\Phi_0\rangle$ are considered. 
In the Fermi state all lowest single-particle levels are occupied up to the Fermi surface,
\begin{equation}
|\Phi_0^{({\rm Fermi})}\rangle = \prod_{k=1}^n p^\dagger_k|0\rangle=|\underbrace{1,1\dots1}_{n\, {\rm spaces}},0,0\dots\rangle.
\label{eq:fermi}
\end{equation}
The Fermi state is an exact ground state of non-interacting  fermions ($G=0$ limit). 

The BCS solution offers a second convenient starting wave function
 $\prod_{k=1}^\omega (u_k+v_k\,p^\dagger_k)|0\rangle.$ In our applications it is projected  
onto an appropriate number of particles; therefore  $|\Phi_0^{({\rm BCS})}\rangle$ is defined via amplitudes as
\begin{equation}
\langle {\bf n}|\Phi_0^{({\rm BCS})}\rangle=\prod_{k=1}^\omega (u_k\delta_{n_k,0}+v_k \delta_{n_k,1}).
\label{eq:BCS0}
\end{equation} 
The coefficients $u_k$ and $v_k$ are determined by solving the usual BCS equations. 
Our procedure does not require the starting wave function to be normalized.
Given a product form of Eq. (\ref{eq:BCS0}), it is efficient to generate  
the BCS based initial state stochastically by selecting $|{\bf n_0}\rangle$ in a process where each randomly selected state $k$ 
is chosen to be occupied or empty with a probability proportional to the corresponding $v_k$ and $u_k;$ the process is stopped once a 
desired number of occupied states given by the total number of particles is reached. 
It is important that variations in implementation of the projected BCS or not following Eq. (\ref{eq:BCS0}) exactly, are not essential 
since the starting wave function can be arbitrary. 

In Fig.~\ref{fig:no_wf_two}(a) the convergence of energy as a function of $L,$ following 
Eq.~(\ref{eq:HLproj}), is shown for the two initial wave functions.  The method just outlined is based on evaluation of the linear norm, the sum of all walkers' bags, and since the actual wave functions are never constructed these are labeled as ``no-wf'' in Fig.~\ref{fig:no_wf_two}. 
In panel (b) the convergence is shown 
as a function of the imaginary time $\tau$ using Eq.~(\ref{eq:expproj}). 
The common energy scale is used in both panels and the energy obtained 
from BCS approach and from the exact diagonalization of the pairing Hamiltonian are shown with horizontal grid lines. 
The magnified 
energy scale used here allows one to clearly see  the difference between BCS and the exact solution.  

As appropriate in a variational technique, 
the BCS ground state energy is above the exact one. 
However, the estimates for the ground state energy, using the linear norm, 
%in Eqs.~(\ref{eq:HLproj}) and (\ref{eq:expproj}) 
approach the exact value from below. This feature, as discussed in Sec.~\ref{sec:error}, is used for providing a lower bound for the 
ground state energy estimate.

%Therefore, 
%in the full variational  CSMC algorithm when the estimate for the ground state wave function is available 
%linear norm energy estimate used in conjunction with a traditional variational estimate $\langle \Phi_L|H|\Phi_L \rangle/\langle \Phi_L|\Phi_L \rangle$ provide lower and upper bounds for the energy estimate. 

In order to compare the projection with power function in Eq.~(\ref{eq:HLproj}), and using the exponential in Eq.~(\ref{eq:expproj}), 
Fig.~\ref{fig:no_wf_two}(b) includes an additional $\overline{L}$ scale shown at the top. The quantity $\overline{L}$ is defined 
as the average number of steps that needs to be taken
by walkers in order for the series (\ref{eq:exptaylor}) to converge for a given imaginary time $\tau.$
While both panels (a) and (b) look similar, using the exponential as a 
projector is more computationally expensive as it requires almost three times as many steps. 

The use of exponent to project a ground state 
does not provide any additional numerical stability; fluctuations at remote times, in cases with no wave function, are
seen in both panels of Fig.~\ref{fig:no_wf_two}. 
These fluctuations are removed by reconstructing wave functions at certain steps; the corresponding curves in Fig.~\ref{fig:no_wf_two} 
are labeled with ``wf''. 
The origin of these fluctuations and error analysis are addressed next. 
Since the exponential projection using imaginary time is deemed to be less effective we will not discuss it any further.

\subsection{Error and convergence control\label{sec:error}}
In the CSMC there are generally two kinds of errors. The first one is the statistical error that emerges as a result of stochastic 
evaluation, for example, estimating wave functions using Eq.~(\ref{eq:mc}) or evaluation of the 
linear norm in Eq.~(\ref{eq:lnorm}). The second error is 
associated with the algorithm used to obtain physical quantities of interest; for example, 
in projection technique this concerns the quality of approximation  
$E_0(L)\approx E_0$  in Eq.~(\ref{eq:HLproj}).  
In this subsection we examine both of these errors and methods of their control. 

The Central Limit Theorem (CLT) is at the core of statistical error control. It is usually expected that as the number of samples, $N,$ grows the 
associated standard deviation $\sigma$  of the ensemble average
% as those in Eqs.~(\ref{eq:mc}) and Eq.~(\ref{eq:lnorm}), 
goes down as 
$\sigma\propto 1/\sqrt{N}.$ However, in CSMC the main disadvantage of independent walks is that 
the variance grows exponentially as the path length increases. 
Therefore,  for large number of steps the average of bags in Eq.~(\ref{eq:lnorm}) 
is hard to evaluate because the distribution of bags becomes too broad. This is
the cause of fluctuations seen in  Fig.~(\ref{fig:no_wf_two}) at large  $L.$

Let us analyze this problem. Consider an ensemble of all $L$-step bags for all possible paths
$\{B_L\},$ let $\sigma^2\{B_L\}$ be its variance and ${\overline B_L}$ its mean.  According to Eq.~(\ref{eq:lnorm}) 
${\overline B_L}=||\Phi_L||.$ 
As proved earlier, all bags are positive making the coefficient of variation ${\rm Cv}\{B_L\}\equiv \sigma\{B_L\}/{\overline B_L}$ an 
appropriate measure of relative error. 
Indeed CLT implies that with $N$ estimates of energy using Eq.~(\ref{eq:HLproj}) the relative error is 
\begin{equation}
\Delta E_0(L) /E_0(L)\approx \frac{1}{\sqrt{N}}  {\rm Cv}\{B_{L+1}\}.
\end{equation}

The problem with divergent behavior of ${\rm Cv}\{B_{L}\}$ as a function of $L$  arises due to $B_L^{(s)}$ for each walker $s$ 
being a product of matrix elements weighted by the corresponding probability, see Eq.~(\ref{eq:bag2}). 
The  product of a large number of random matrix elements is poorly behaved. 
Let us assume that $c$ gives the coefficient of variation for all possible matrix elements weighted by chosen probabilities,
then each term in the product (\ref{eq:bag2}) has a coefficient of variation,
\begin{equation}
{\rm Cv}\left \{ \frac{\langle {\bf n}_{L} | V | {\bf n}_{L-1} \rangle}{{\cal P}\left({\bf n}_{L-1} \rightarrow {\bf n}_{L} \right)}  \right \} \equiv c^2.
\end{equation}
Then the standard statistical treatment of a product leads to  
\begin{equation}
{\rm Cv}\{B_L\}= \sqrt{(1+c^2)^L-1 };
\label{eq:staterr}
\end{equation}
this simple form is obtained under the assumption of uniform initial distribution in Eq.~(\ref{eq:bag1}).
With the exemption of some special cases, $c>0;$ in most situations of interest $c\approx 0.5,$ therefore ${\rm Cv}\{B_L\}$ grows exponentially 
with $L.$ 
Thus, it is practically impossible to 
compete with the exponentially increasing variance of the distribution by increasing 
the number of walkers.
%Eq.~(\ref{eq:staterr}) shows that in order to maintain a reasonable level of error under typical numerical limitations 
%walkers should not be allowed to walk independently for more than 20-60 steps. 

\begin{figure}[!h]
\centering
\includegraphics[width=80mm]{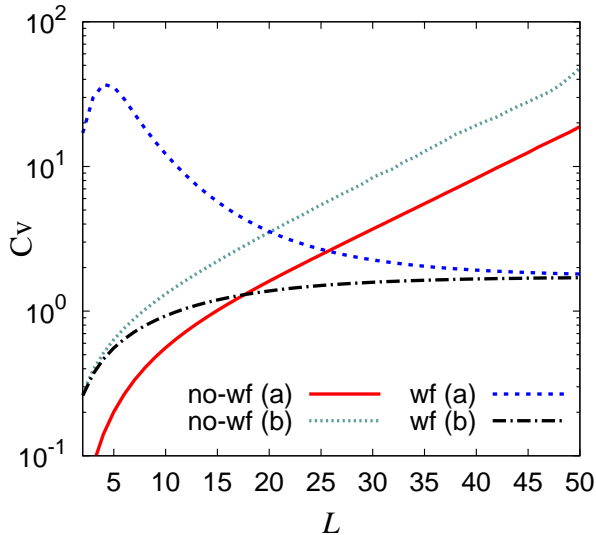}
\caption{(Color online) The half-occupied ladder model with $n=9,$ $\omega=18,$ and $G=1.$ 
Coefficient of variation  ${\rm Cv}\{B_L\}$ is shown as a function of $L$.  The figure includes four curves with two 
possible initial states:
(a) Fermi state $\Phi_0^{({\rm Fermi})}$ Eq.~(\ref{eq:fermi}) and (b) constant-component state 
where $\langle {\bf n}|\Phi_0^{({\rm const})}\rangle=1;$ and for calculations with and without wave functions being obtained. 
}
\label{fig:cv_plot}
\end{figure}

In Fig.~(\ref{fig:cv_plot}) the behavior of  
${\rm Cv}\{B_L\}$ as a function of $L$ is shown for the half-occupied 18-level ladder model.  
The results for two different starting 
states are shown. 
The exponential divergence
in Eq.~(\ref{eq:staterr}) pertains to the situation involving independent walkers, where the actual wave function is not obtained.
The two corresponding curves, labeled with ``no-wf'',   
both display the same 
exponential divergence with $c\approx 0.42$ which corresponds to asymptotic behavior, ${\rm Cv}\{B_L\}\propto 1.086^L.$

The limitation on the number of independent steps is relatively easy to overcome. 
The exponential growth of ${\rm Cv}\{B_L\}$ is usually weak, and in most cases, such as the 
example in Fig.~\ref{fig:no_wf_two}, no problems emerge for  
$L$ less than 30 or 50. Moreover,  the
choice of probabilities that follows importance sampling in Eq.~(\ref{eq:importance}) would lead to $c=0.$ 
Practically, numerical noise never allows one to reach this ideal limit but the 
the growth of variance can be delayed.
In addition to that, 
with a good initial wave function, such as the one from BCS theory,
the convergence is reached in a few steps, before the onset of statistical problems; see example in Fig.~\ref{fig:no_wf_two}.

Preventing walkers from taking long independent walks,
by combining them in wave functions after a certain number of steps with Eq.~(\ref{eq:mc}),
allows one to  avoid the problem completely. This is demonstrated in Fig.~\ref{fig:no_wf_two} with curves labeled ``wf''.
The intermediate summation at a moment when bags are combined, due to the CLT, prevents an exponential increase of the variance.   
In the implementation of the CSMC 
the statistical error is tracked by controlling the coefficient of variation in the bags as $L$ is increased. This allows one to apply a computationally expensive procedure of reconstructing the wave function only when necessary, typically once in every 5--20 steps. 
At a moment when the full wave function is built, the convergence of the projection technique 
(the second kind of error) can be assessed using the usual square, ${\cal L}_2,$ norm.

The second kind of errors, which is convergence $E_0(L)\rightarrow E_0$  in these examples, is a part  
of any iterative diagonalization technique, such as Lanczos or Davidson algorithms; and it has been well 
studied in the past. However, the specifics of the pairing problem described in Sec.~\ref{sec:features} and the use of the linear norm allow one to place exact upper and lower limits on the value of energy. 

\begin{figure}[!h]
\centering
\includegraphics[width=80mm]{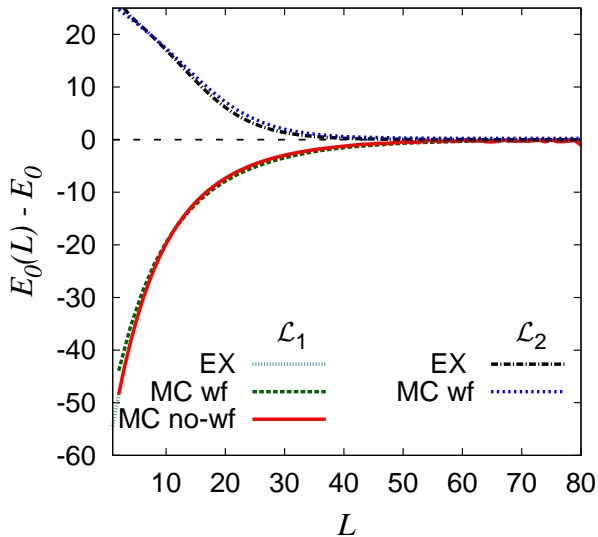}
\caption{(Color online) The half-occupied ladder model with $n=9,$ $\omega=18,$ and $G=1.$  Deviation of the energy estimates using linear  ${\cal L}_1$ and square ${\cal L}_2$ norms from Eqs.~(\ref{eq:HLproj}) and (\ref{eq:l2norm}), respectively, is shown as a function of $L.$ The curves correspond to exact, CSMC with and without wave function reconstruction.  
}
\label{fig:error}
\end{figure}

The convergence of the projection 
algorithm is examined in Fig.~\ref{fig:error}. Here, using the same half-occupied ladder model with $\omega=18,$ we show deviation of the predicted energy from the exact value as a function of the number of steps. Three curves, that are essentially indistinct, show $E_0(L)-E_0$ where 
$E_0(L)$ was evaluated using the linear norm ${\cal L}_1,$ Eq.~(\ref{eq:HLproj}). The three sets of results are obtained 
by evaluating  $|\Phi_L\rangle$
exactly with matrix-vector multiplication (dotted line); using CSMC with wave function being reconstructed at each step (dashed line); and without 
the wave function using bag average in Eq.~(\ref{eq:lnorm}) (solid line).  
%For the latter case at high $L$ one can notice a contribution from statistical fluctuations, which we have just discussed.  
The slight difference between exact and CSMC results is only due to an intermediate shift by chemical potential.  
These three curves approach the exact energy from below, which is a distinct property of the ${\cal L}_1$ norm. 

The other two, nearly indistinct, curves show the convergence of energy evaluated using the traditional square norm, labeled as ${\cal L}_2$
\begin{equation}
E_0(L)=\frac{\langle\Phi_L|H|\Phi_L\rangle}{\langle\Phi_L|\Phi_L\rangle}.
\label{eq:l2norm}
\end{equation}
The ${\cal L}_2$ norm can be used only  when the wave function is available, for that reason only the curves for exact (dash-dot) and CSMC with wave function (short dash) appear in Fig.~\ref{fig:error}. Naturally, the expectation value of the Hamiltonian in Eq.~(\ref{eq:l2norm}) is subject to the variational principle, and all curves with ${\cal L}_2$ norm approach ground state energy from above. Thus, the estimates using
${\cal L}_1,$ Eq.~(\ref{eq:HLproj}), and ${\cal L}_2$,  Eq.~(\ref{eq:l2norm}) norms give the lower and upper bounds for the value of ground state energy. 

To summarize, in our algorithm we rely on computationally inexpensive independent propagation of walkers in configuration space until 
the coefficient of variation of their bags exceeds some critical value. At that moment the full wave function is reconstructed and is used to 
evaluate energy from Eq.~(\ref{eq:l2norm}) and all other operators of interest. The combination of energy estimates from linear and square 
norms give lower and upper bounds for the actual value of energy. If the desired convergence is not reached, the process is 
continued starting from the current wave function. 

\subsection{Weak pairing limit}
\begin{figure}[!h]
\centering
\includegraphics[width=80mm]{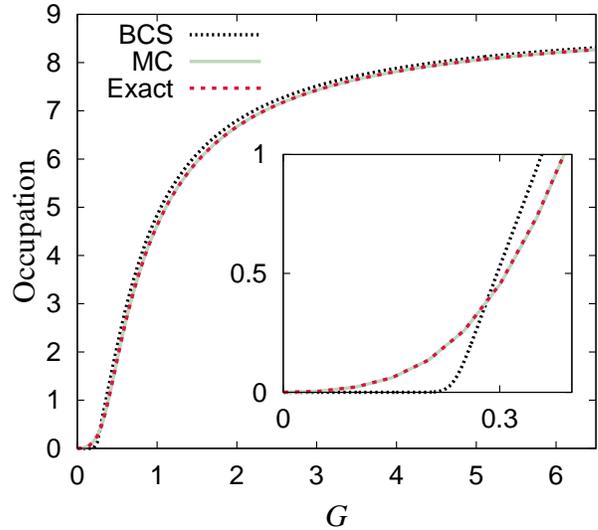}
\caption{(Color online) The half-occupied ladder model with $\omega=18$ and $G=1$. The net occupancy (total number of particles) on the upper 9 orbitals is 
shown as a function of the pairing strength. The weak pairing limit is magnified in inset. The three curves correspond to BCS solution, CSMC (MC) solution, and the exact solution by means of diagonalization. The CSMC and exact results are indistinct and the corresponding curves are overlaid. 
For the CSMC solution we used $N=7.5\times 10^5$ walkers, limiting the number of independent steps to five. 
}
\label{fig:gcrit}
\end{figure}

As mentioned earlier,  superconducting paired states in small systems face a lot of competition from other incoherent interactions as well as 
from the single particle shell structure.  
Thus, relatively weak and fragile superconducting states is one of the distinct characteristics of pairing in nuclei. 
Unfortunately, the BCS theory is not designed to work in this limit, and having the 
CSMC as a computationally inexpensive alternative is one of the main
motivations of this work.  In Fig.~\ref{fig:gcrit} we demonstrate the effectiveness of CSMC in the limit of weak pairing 
using our half-filled 18-level ladder model. In the limit when the pairing strength $G=0,$ the system settles in the Fermi state with 9 
lowest double-degenerate single-particle states being occupied. 
As soon as $G>0,$ pair excitation promotes particles up, and the occupation of the upper 9 levels 
becomes non-zero. For very strong pairing, $G\gg\epsilon,$ the limit of 
degenerate model with equal occupancy of all states is reached. This limit leads to half of the 18 particles being on lower 9 levels and
half on the upper ones.  

In Fig.~\ref{fig:gcrit} the net occupation of the upper 9 levels as a function of G is shown. This plot includes results from 
BCS, CSMC (labeled as ``MC'') and 
exact diagonalization.  While on a large scale all results are similar, in the region of low pairing strength, which is shown in inset, 
the well known problem with BCS solution, shown in dotted (black) line, is noticeable. 
At the same time, exact and CSMC results are indistinct; dashed (crimson color) line goes 
right on top of the solid (sea-green color) line.  
This test illustrates that the CSMC is well suited for all limits of pairing strength. Moreover, in the limit of weak pairing, the computational effort in 
CSMC is reduced as the contribution from rare excursions above Fermi surface can be easily evaluated with importance sampling.

\subsection{Excited states\label{sec:ex}}
Obtaining excited states with CSMC is more computationally difficult. One can no longer use a linear norm since all amplitudes cannot 
be positive definite simultaneously; that is, statement (vi) in Sec.\ref{sec:features} is not valid for excited states. 
Therefore, the usual quadratic ${\cal L}_2$ norm has to be used and the bag values can be negative. 
Nevertheless, features (i)-(v) in Sec.\ref{sec:features} remain valid and useful. In particular,  
since the matrix elements of $V$ are positive definite, the importance sampling is still an effective strategy and 
the signs of bags are not altered by repeated application of the 
Hamiltonian which curtails the typical MC sign problem. 
Similar, to Lanczos technique, the CSMC approach requires orthogonalization, therefore the wave functions have to be built each time the orthogonalization is to be performed. The need for orthogonalization limits the number of independent steps, which is the main reason 
for higher computational demand. 

\begin{figure}[!h]
\centering
\includegraphics[width=80mm]{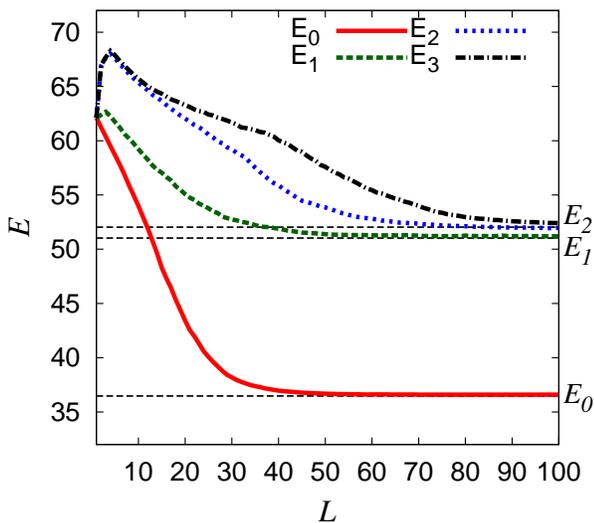}
\caption{(Color online) The half-occupied ladder model with $\omega=18$ and $G=1$.
Convergence of CSMC to ground state, to first excited state and to double-degenerate second excited state is shown. %{\red What about 4-th line to degenerate state?}
}
\label{fig:excited18}
\end{figure}

In Fig.~\ref{fig:excited18} we show the CSMC applied to the study of excited states in the same half-occupied 18-level ladder model. 
The ladder model example is particularly challenging since the density of states above the gap is high. In this model the 
level spacing between the ground and first excited state, which is about twice the BCS gap, is 
$E_1-E_0 \approx 14.6$ (in units of  level spacing $\epsilon=1$). At the same time, the spacing between the following states $E_2-E_1 \approx 0.3$ 
is very small. Moreover, the second excited state is double-degenerate. 

%\section{Examples of applications\label{sec:examples}}

\subsection{Pairing in Sn isotopes}
In order to illustrate the CSMC algorithm in a realistic case where pairing matrix elements are not all equal to a constant, we consider isotopes of tin. 
The role of pairing in $^{100-132}$Sn isotopes has been extensively explored in the literature 
\cite{Dean:2002zx,Brown:2002,Zelevinsky:2003,Chen:2014}.
Apart from questions of scientific interest such as pairing matrix elements and their connection to superconducting state in infinite matter, near
constancy of the excitation energy of the lowest $2^+$ states, and unexplained behavior of electric quadruple transition rates, the tin case
emerged as a benchmark for computational techniques.  In Tab.~\ref{tab:tin} we present comparison of energies and occupation numbers for 
$^{116}$Sn, $^{118}$Sn, and $^{120}$Sn. The model space here includes five single-particle levels (with total $\omega=16$), their energies and spins are listed in the 
first two columns of Tab.~\ref{tab:tin}, the matrix elements are taken from 
from the G-matrix calculation in Ref.~\cite{Holt:1998}, the values can be found in Table~1 of Ref.\,\cite{Zelevinsky:2003}.
The results in Tab.~\ref{tab:tin} show expected level of agreement. 
With increased computational effort, mainly using large number of walkers, any desired
level of precision can be obtained; our goal here was to use minimal effort and to solve the pairing problem with a precision that exceeds 
any practical need, which is set to be 5 keV uncertainty for energy and $0.01$ for occupation numbers. 
\begin{table*}
    \begin{tabular}{|  c  c |  c  c  | c  c | c  c |}
    \hline \hline
     &  &$^{116}$Sn &  & $^{118}$Sn &  & $^{120}$Sn & \\ 
        &  & Exact & CSMC & Exact & CSMC & Exact & CSMC\\ \hline \hline
    j&$E_0$(MeV)   & -153.766 & -153.765 & -170.115 & -170.113 & -185.945 & -185.945  \\ \hline\hline
%     $j$ & $\epsilon_{sp}$ & $Exact$ & $MC$ & $Exact$ & $MC$ & $Exact$ & $MC$\\ \hline
     ${5}/{2}$ & $-9.736$ & 4.99 & 4.99 & 5.13 & 5.13 & 5.25 & 5.25 \\ \hline
     ${7}/{2}$ & $-8.957$ &  5.88 & 5.89 & 6.28 & 6.27 & 6.60 & 6.60 \\ \hline
     ${1}/{2}$ & $-7.302$ & 0.67  & 0.67 & 0.76 & 0.76 & 0.86 & 0.86 \\ \hline
     ${3}/{2}$ & $-7.634$ &  1.18  & 1.17 & 1.63 & 1.63 & 2.08 & 2.09 \\ \hline
     ${11}/{2}$ & $-7.544$ & 3.29 & 3.28 & 4.21 & 4.20 & 5.21 & 5.21 \\ \hline
    \end{tabular}
\caption{Comparison of exact and CSMC results for selected isotopes of tin. After header, first row shows comparison of energies, 
the remaining five rows show occupation numbers for five singe particle states.   The calculation of energies is done with 
$N=5\times 10^6$ walkers using linear norm. The final error is about 5 keV.
\label{tab:tin}}
\end{table*}

\subsection{Large scale model}
As a final illustration of the CSMC algorithm we explore a model of the $^{24}$O nucleus 
intended to reflect the nature of  pairing correlations 
in a system containing both bound states and a continuum of scattering states. 
Our main goal is to demonstrate the capabilities of our algorithm 
while addressing the problem of pairing in continuum qualitatively. 
Quantitative studies require good knowledge of the effective interaction Hamiltonian; construction of this Hamiltonian is outside the scope of this presentation.  

For our study we select the Woods-Saxon potential with parameters from Ref. \cite{Schwierz:2007} to model the mean field of weakly-bound $^{24}$O nucleus. We discretize this potential using a large quantization-box of size 500 fm. This allows us to generate a dense continuum of states. We limit scattering states by about 8 MeV of energy, which leads to $\omega$ of about 100. 
For the pairing interaction between neutrons we use a density dependent contact interaction from 
Refs.~\cite{Bonche:1985,Chasman:1976, Dobaczewski:1996} 
\begin{equation}
V(\textbf{r}, \textbf{r}') = -G_{0}\left (1 - \eta \frac{\rho(\textbf{r})}{\rho_{0}}\right ) \delta(\textbf{r} -  \textbf{r}').
\end{equation}
Here $\rho(\textbf{r})/\rho_{0}$ is the nucleonic density expressed relative to the saturation density. 
This quantity is assumed to be given by the Woods-Saxon form factor. 
The density dependence of pairing is controlled by a parameter $\eta$ which is selected as $\eta = 0.5.$ Following 
Ref.~\cite{Bonche:1985,Dobaczewski:1996}, we also introduce a momentum cut-off function, that gradually  
reduces the pairing matrix elements to zero for scattering states at energies above 5 MeV, 
the diffuseness parameter of the cut-off function is 0.5 MeV, see also Ref.~\cite{Lingle:2015}.

%Following the approach put forth in \cite{Dobaczewski:1996} and \cite{Bonche:1985}, the antisymmetric pairing matrix elements $G_{kk'}$ are defined to be
%\begin{equation}
%G_{kk'} \equiv f_{k} ( \langle k \tilde{k} \left| G \right| k' \tilde{k'} \rangle - \langle k \tilde{k} \left| G \right| \tilde{k'} k' \rangle ) f_{k'}
%\end{equation}
%where the $f_{k}$ are energy cutoff factors defined as
%\begin{equation}
%f_{k} = \left[ \frac{1}{1 + e^{(\epsilon - a)/b}} \frac{1}{1 + e^{(-\epsilon - a)/b}} \right]^{1/2}
%\end{equation}
%with $a$ set to the energy of the lowest continuum level and $b = 0.5$ MeV. These factors are included to subdue the unrealistic pair transfer into highly unbound states. 

For our example we assume an inert $^{16}$O core which leaves two bound $s_{1/2}$ and $d_{5/2}$ valence single-particle states. 
Therefore, the bound states can accommodate $n = 4$ pairs of valence neutrons in  $^{24}$O.
The pairing matrix elements involving these states are known from the phenomenological shell model Hamiltonian in Ref. \cite{Brown:2006}. 
Following previous studies, we adopt the value of $G_{0} = 1$ GeV$\cdot$fm$^3$ for treating pairing interaction involving the continuum of scattered states. This value is also consistent with the pairing strength in phenomenological Hamiltonians \cite{Brown:2006}. 
Here we limit our consideration to $s$-wave single-particle continuum. Due to centrifugal barrier the overlap between bound and unbound $d$-wave states is small; this inhibits virtual pair excitations to $d$-wave states in the continuum. 

\begin{figure}[!h]
\centering
\includegraphics[width=80mm]{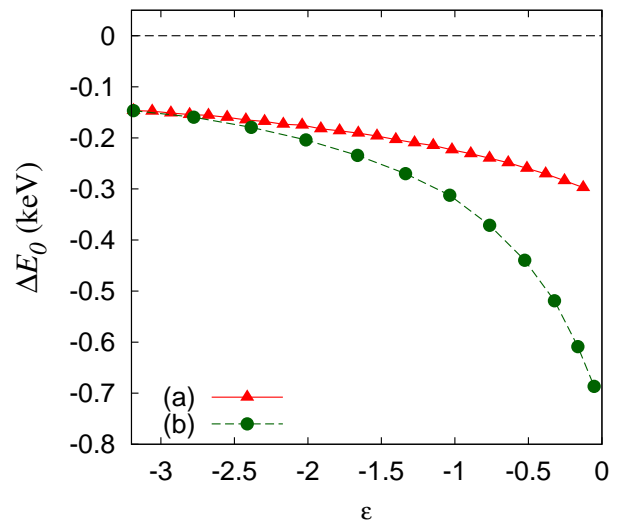}
\caption{(Color online) The energy correction due to the inclusion of continuum states for a pairing model with $\omega = 100$ and $n = 4$. The correction, $\Delta E_0$, 
is the difference between the CSMC result with continuum states and an exact answer for a model including only the two bound states. The $
\epsilon$ along the lower axis is the energy of the s-wave bound state as it is moved closer to the continuum. The Monte Carlo error is negligible.
}
\label{fig:cont_correction}
\end{figure}

Our goal in this investigation is to estimate the role of continuum. We do this by comparing full calculation with the one 
where the continuum is ignored.  In Fig.~\ref{fig:cont_correction} the change in the ground state energy $\Delta E_0$  
is shown as a function of the single-particle energy $\epsilon$ of the $s_{1/2}$
state.  We present two different cases.  In case {\sl (a)}  the parameters of the pairing Hamiltonian, which includes the 
single particle energies and pairing matrix elements, are first evaluated with a realistic choice of 
Woods-Saxon parameterization for $^{24}$O and then the $\epsilon$ which corresponds to $s_{1/2}$ state,
is varied while all other parameters remain unchanged. In the self-consistent case {\sl (b)} the depth of the Woods-Saxon potential is varied which moves the  $s_{1/2}$ state, and each time a new configuration space Hamiltonian matrix is calculated and studied.   

Let us summarize this study. First, the correction from pair excitations into continuum appears to be relatively small, here 
it is of the order of one kilovolt, 
for all reasonable choices of pairing strength $G_0$ the effect is not expected to exceed a few tens of kilovolts, see Ref.~\cite{Lingle:2015}. 
The smallness of the effect does not seem to contradict observations. So far there has been no significant near-threshold discontinuity observed 
in nuclear structure that can be attributed to two-body decay or to pair excitations. 
The decay of $^{26}$O is observed to be very slow, see discussion \cite{Volya:2014b},
which through dispersion relations indicates weakness of the continuum coupling. 

Second, as expected, the effect increases sharply as the bound state approaches the continuum threshold. 
This is similar to the results known for single particle states, while the exact 
near threshold behavior is defined by the phase space volume, see Refs.~\cite{Volya:2012nsrt,Volya:2014}. 

Third, the difference between the two models highlights the importance of halo phenomenon and its proper treatment. In model {\sl (b)}
the wave function of the single-particle $s_{1/2}$ state spatially extends as its energy approaches the threshold. This facilitates pairing 
in the continuum and the resulting effect is significantly stronger than that in model {\sl (a)} where the spatial structure of the single particle wave function 
was not modified. 

Finally, we find that this example successfully demonstrates the power of the CSMC method.

%The Monte Carlo parameters for this model were $N = 6 \times 10^{7}$ walkers, $L$ = 300, and the number of independent steps was limited to one.

%Additionally, figures (\ref{fig:cont_correction_shiftv}) and (\ref{fig:cont_occupation_shiftv}) show the correction 
%obtained by varying the Wood-Saxon potential depth so that the s-wave bound state is moved closer to the continuum. 
%The Monte Carlo parameters are the same as before: $N = 6 \times 10^{7}$ walkers, $L$ = 300, and the number of 
%independent steps was limited to one.

\section{Summary}
In this work we put forward a new Configuration Space Monte Carlo method for solving the many-body pairing problem in finite systems. 
Unlike previous Monte Carlo techniques that deal with pairing interaction in a way similar to MC methods in physics of spin systems,  
our approach does not use evolution in imaginary time, does not need Trotter-Suzuki propagator breakup, and does not depend on 
the pairing matrix elements being constant. 
We propose to evaluate Hamiltonian and other 
observables by stochastically evaluating the corresponding operators in the Krylov subspace spanned by 
states formed as a result of 
powers of Hamiltonian acting on an arbitrary initial state. States in the Krylov subspace are evaluated using random walks in 
the many-body configuration space.  Importance sampling is used to effectively probe components of the wave functions. 
We emphasize several important features of the pairing Hamiltonian, that make the MC approach appealing. In particular, we stress 
boson-like behavior of nucleon pairs, absence of the fermion sign problem, potential for probabilistic interpretation
of transitions in configuration space, and probabilistic interpretation of ground state amplitudes. 
In addition to traditional quadratic quantum mechanical norm, probabilistic 
interpretation allows us to use a linear norm. 
We demonstrate that the approach based on the linear norm is computationally efficient, is perfect for parallelization,  and provides 
effective methods for control of errors of both stochastic and non-stochastic origins.   

The workings of the CSMC method are demonstrated with several examples. 
With a classic ladder model we demonstrate convergence using several variations of the method; 
we discuss errors and present effective means of their control.
The effectiveness of CSMC in small systems where pairing can be effectively weak is shown; the CSMC in its most complete form is used 
for obtaining degenerate and nearly-degenerate excited states in the ladder model. 

As a realistic example, we use isotopes of tin which represents another well studied classic case of pairing in nuclei. The energies and occupation 
numbers in this non-constant pairing example are consistent with exact results.   
Large-scale study of pairing correlations is illustrated using a model of $^{24}$O that includes a continuum of scattering states. 
While our last example is still far from realistic, it highlights the effectiveness of CSMC, and suggests an arena for future applications of the method.

\begin{acknowledgments}
This material is
based upon work supported by the U.S. Department of Energy Office of Science, Office of
Nuclear Physics under Award Number DE-SC0009883.
\end{acknowledgments}

%\bibliography{mcpairing}

\end{document}